\documentclass[aps,prb,twocolumn, superscriptaddress,amsmath,showpacs, showkeys]{revtex4}

\usepackage[latin1]{inputenc}
\usepackage{graphicx}
\usepackage{dcolumn}
\usepackage{bm}
\usepackage{setspace}

\newcommand {\dg} {\ensuremath{^{\circ}}}
\newcommand {\mub} {\ensuremath{\mu_{\text{B}}}}

\newcommand {\V}[1] {$\mathbf{#1}$} 
\newcommand {\etal} {\textit{et al.}}
\newcommand {\ICMu} {IC$\mu$}
\newcommand {\TN} {\ensuremath{T_{N}}}
\newcommand {\TQ} {\ensuremath{T_{Q}}}
\newcommand {\TC} {\ensuremath{T_{C}}}
\newcommand {\TIII} {\ensuremath{T_{\mathrm{III}}}}
\newcommand {\CePrB}[2] {Ce$_{#1}$Pr$_{#2}$B$_6$}
\newcommand {\CeLaB}[2] {Ce$_{#1}$La$_{#2}$B$_6$}

\newcommand {\CePrBx} {\CePrB{x}{1-x}}
\newcommand {\CeLaBx} {\CeLaB{x}{1-x}}

\begin{document}

\title{Magnetic order and multipole interactions in \CePrBx\ solid solutions}




\author{J.-M. Mignot}
\author{G. Andr\'{e}}
\author{J. Robert}
\affiliation{Laboratoire L\'{e}on Brillouin, CEA-CNRS, CEA/Saclay, 91191 Gif sur Yvette (France)}

\author{M. Sera}
\author{F. Iga}
\affiliation{Department of Quantum Matter, ADSM, Hiroshima University, Higashi-Hiroshima, 739-8530 (Japan)}

\date{\today}


\begin{abstract}
Magnetic ordering phenomena in \CePrBx\ solid solutions have been studied using both powder and single-crystal neutron diffraction. A variety of magnetic structures are observed depending on temperature and Ce concentration. Over a broad composition range ($x \le 0.7$), Pr--Pr interactions  play a dominant role, giving rise to incommensurate structures with wave vectors of the form $\mathbf{k}_{\mathrm{IC1}}=(1/4-\delta, 1/4, 1/2)$ or $\mathbf{k}_{\mathrm{IC2}}=(1/4-\delta, 1/4-\delta, 1/2)$. The crossover to a \hbox{CeB$_{6}$-like} regime takes place near $x = 0.7$--0.8. For the latter composition, the antiferroquadrupolar phase transition observed in transport measurements precedes the onset, at lower temperature, of a commensurate magnetic order similar to that existing in CeB$_{6}$. However, unlike in the pure compound, an incommensurate magnetic order is formed at even higher temperature and persists in the antiferroquadrupolar phase down to the lock-in transition. These results are shown to reflect the interplay between various type of dipole exchange and higher multipole interactions in this series of compounds.
\end{abstract}




\maketitle


\section{\label{sec:intro}Introduction}

Light rare-earth hexaborides (\textit{R}B$_{6}$, \textit{R} = Ce, Pr, Nd) have attracted much interest for their unconventional ordering properties originating from different types of multipole interactions. They crystallize in a simple-cubic, CsCl-type structure (space group no.\ 221, $Pm\bar{3}m$), which is favorable for retaining high degeneracies and orbital degrees of freedom, within the crystal-field eigenstates.  The dense Kondo compound CeB$_{6}$ is well known for developing antiferroquadrupolar (AFQ) long-range order below a transition temperature $\TQ= 3.2$ K (phase II). This order consists of a staggered arrangement, with wave vector $\mathbf{k}_{\mathrm{AFQ}}=(1/2, 1/2, 1/2)$, of $O_{xy}$-type quadrupole moments ($O_{xy}=[\sqrt{3}/2]\;[J_{x}J_{y}+J_{y}J_{x}]$) associated with the $\Gamma_{8}$ quartet ground state of Ce$^{3+}$, as was first established by Effantin \etal\ using neutron diffraction.\cite{Effantin'85} The latter work represented a milestone in multipole interaction studies because it showed that, although quadrupole order does not break time-reversal symmetry and thus cannot be directly observed in a neutron experiment, it can produce an antiferromagnetic (AFM) dipole component at the same wave vector $\mathbf{k}_{\mathrm{AFQ}}$, which is easily detectable, under the application of a magnetic field. Subsequently, nonresonant x-ray diffraction (XRD) measurements\cite{Tanaka'04} confirmed this AFQ order, accompanied by a higher-rank contribution from electric hexadecapole moments. On the theoretical side, Sakai et al.,\cite{Sakai'97} have emphasized the central role played by magnetic octupole interactions in the properties of phase II and shown, in particular, that the extra hyperfine field produced by octupole moments could solve a longstanding inconsistency between neutron and NMR results in this material. 

Below $\TIII=2.3$ K, CeB$_{6}$ undergoes a second transition into a long-range magnetic ordered state (phase III). Its magnetic structure,\cite{Effantin'85} denoted $2k$-$k'$, is non-collinear and involves four different Fourier components corresponding to the commensurate wave vectors, $\mathbf{k}_{C}^{(1,2)}=(1/4, \pm 1/4, 1/2)$ and $\mathbf{k}_{C}'^{(1,2)}=(1/4, \pm 1/4, 0)$. It can be described as a stacking of \{001\} planes in which the Ce moments are alternatingly oriented along two orthogonal (in-plane) binary axes.\footnote{Alternative solutions have been proposed subsequently in Refs. \onlinecite{Zaharko'03} and \onlinecite{Iwasa'03}. Their main features ($2k$--$k'$, planar, transverse sine-wave character, with orthogonal moments at neighboring sites directed along \{110\} axes) are the same as in Effantin's model, and the differences are not essential for the arguments discussed in this report.}

 \begin{figure} [b]
	\includegraphics [width=0.95\columnwidth, angle=0] {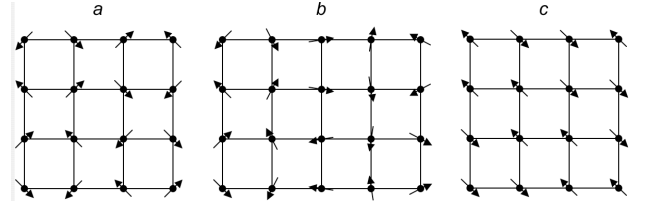}
\caption{\label{PlanarStruc} Schematic representation of planar magnetic structures proposed for CeB$_{6}$ and PrB$_{6}$ in Refs.~\onlinecite{Effantin'85} and \onlinecite{Burlet'88}: (a) commensurate, double-$k$, (b) incommensurate,  double-$k$, (c) commensurate,  single-$k$. In the case of PrB$_{6}$, the moments have opposite directions in neighboring planes, whereas they are turned by 90 degrees in phase III of CeB$_{6}$.}
 \end{figure}

It has been argued, both on experimental and theoretical grounds,\cite{Effantin'85,Sakai'99} that this unique structure is closely related to the underlying AFQ order. In this phase too, octupole moments ($T_{xyz}=[\sqrt{15}/6] \; \overline{J_{x}J_{y}J_{z}}$, where the bar denotes the sum of all possible permutations of indices) are thought to play a major role.\cite{Sakai'99} Interestingly, experimental evidence was recently reported for a magnetic octupole order occurring in the as yet elusive ``phase IV'' of \CeLaBx\ ($x = 0.70$--0.75) solid solutions.\cite{Kuwahara'07}.

PrB$_{6}$ presents a contrasting case, in which the $\Gamma_{3}$ (triplet) crystal field ground state of Pr$^{3+}$ can also sustain a quadrupole (QP) moment, but no octupole moment. Long-range AFQ order of the type found in phase II of CeB$_{6}$ does not develop in the Pr compound either at $H=0$ or in an applied field. On the other hand, two magnetic ordered phases exist below $\TN=7$ K and $\TC=4.2$ K, respectively.\cite{McCarthy'80,Effantin'85t,Burlet'88} The lower one is commensurate (C), double-$k$, and associated with the same wave vectors $\mathbf{k}_{C}^{(1,2)}=(1/4, \pm 1/4, 1/2)$ as phase III in CeB$_{6}$, but without the additional $\mathbf{k}_{C}'^{(1,2)}$ components. The magnetic structure existing between \TC\ and \TN\ is also double-$k$ but the wave vectors are now incommensurate (IC), $\mathbf{k}_{\mathrm{IC}}^{(1,2)}=(1/4-\delta, \pm 1/4, 1/2)$ with $\delta=0.05$ (see Fig.~\ref{PlanarStruc}). The lower transition thus corresponds to a lock-in mechanism.\cite{Rossat'87} The similarity with CeB$_{6}$ strongly suggests that the commensurate magnetic order may again be associated with an AFQ order of $O_{xy}$ moments, but the AFM stacking along the direction normal to the planes would entail a different AFQ wave vector $\mathbf{k}_{\mathrm{AFQ}}=(1/2, 1/2, 0)$ to achieve proper matching of the dipole and quadrupole structures. On the other hand, Kuramoto and Kubo \cite{Kuramoto'02} have argued that GdB$_{6}$, which has no 4$f$ orbital degeneracy, orders magnetically with the same wave vector $(1/4, 1/4, 1/2)$ and that, consequently, this wave vector must reflect peculiarities of the Fermi surface which are common to all light rare-earth hexaborides. The physical origin of the complex magnetic structures found in CeB$_{6}$ and PrB$_{6}$ therefore remains controversial.

In order to shed light on this problem, we have prepared \CePrBx\ solid solutions with different concentrations and studied the evolution of their magnetic ordering properties using both powder and single-crystal neutron diffraction. Magnetization and transport measurements were reported previously for this series.\cite{Kim'06,Kishimoto'05} The results show that substituting Pr for Ce in CeB$_{6}$ initially produces a steep decrease in the AFQ transition temperature \TQ, while enhancing \TIII. At the particular composition $x = 0.7$, the initial situation, $\TN < \TQ$, becomes reversed with $\TN=4.1$ K and \TQ\ below 2 K at $H = 0$, but distinct anomalies observed in the physical properties indicate that phase II reappears at high magnetic fields, and also that the $\TQ(H)$ transition line extends well into the magnetic order region at low fields.\cite{Kishimoto'05} The results also reveal the existence of two different magnetic ordered phases at $H = 0$, IC2 forming at $T_{\mathrm{IC2}} \equiv \TN$ and IC1 below a first-order phase transition at $T_{\mathrm{IC1}}=3.5$ K. Preliminary neutron measurements \cite{Mignot'06} showed that phases IC1 and IC2 both have incommensurate magnetic structures. Meanwhile, Tanaka \etal \cite{Tanaka'06} performed nonresonant synchrotron XRD experiments on the same material and observed IC superstructure peaks in phase IC1, which they ascribed to an IC multipole order.

In this paper, we present neutron diffraction results for different compositions, both on the Pr-rich and Ce-rich sides, and discuss how magnetic properties observed by neutrons can be traced back to the effects of Ce and Pr multipole moments. The report focuses on measurements performed at $H = 0$. High magnetic field properties and $(H,T)$ phase diagrams will be the subject of a forthcoming paper.


\section{\label{sec:exp}Experiments} 

\CePrBx\ solid solutions with $x = 0.2$, 0.4, and 0.8 were, prepared in Hiroshima using 99.5\%\ enriched $^{11}$B isotope. Neutron powder diffraction (NPD) patterns were collected at the LLB in Saclay, using the two-axis diffractometer G4-1 (800-cell position-sensitive detector) at an incident wavelength of 0.24266 nm ($x=0.8$) or 0.24226 nm ($x=0.2$ and 0.4). A pyrolytic graphite filter was placed in the incident beam to suppress higher-order contamination. The sample powder was put in a cylinder-shape vanadium container, 6 mm in diameter. The data analysis was performed using the Rietveld refinement program \textsc{FullProf},\cite{fullprof'93,fullprof'01} with neutron scattering lengths and magnetic form factors taken from Refs.~\onlinecite{Sears'92} and \onlinecite{Freeman'79}, respectively. Absorption corrections were applied using an estimated absorption coefficient $\mu R = 0.3$.

\begin{table}  [b] 
\caption{\label{crysparam}Refined nuclear structure parameters of \CePrBx.}
\begin{ruledtabular}
\begin{tabular}{c c c c c c c}
Ce concentration & $T$ (K) & $a$ (\AA) & $x_B$\\
0.2 & 9.9 & 4.1416(2) & 0.1984(5) \\
0.4 & 10.0 & 4.1435(2) & 0.1990(5) \\
0.8 & 8.0 & 4.1484(1) & 0.1985(3)\\
\end{tabular}
\end{ruledtabular}
\end{table}

\begin{figure}  
	\includegraphics [width=0.65\columnwidth, angle=-90] {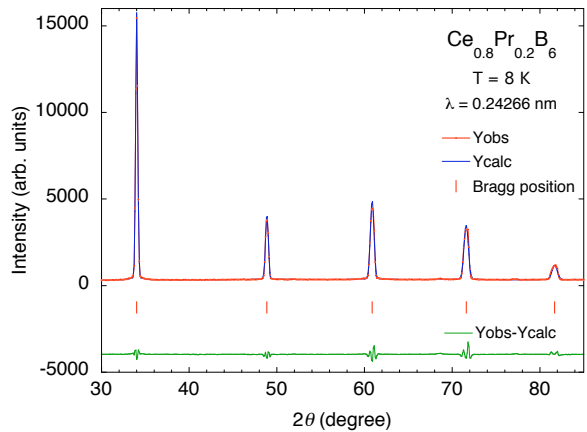}
\caption{\label{Ce0.8nuc} (Color online) Refinement of the neutron diffraction pattern of \CePrB{0.8}{0.2}\ measured in the paramagnetic phase at $T = 8$ K (Bragg reliability factor $R_{B}=0.017$).}
 \end{figure}

\begin{figure}  
	\includegraphics [width=0.50\columnwidth, angle=-90] {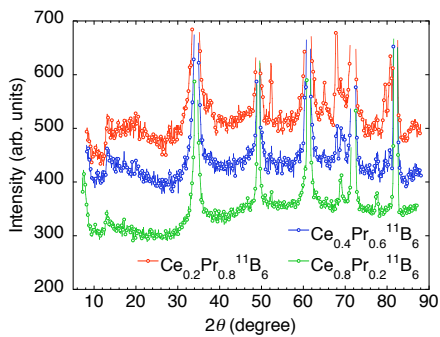}
\caption{\label{paraphase} (Color online) Expanded plot of the neutron diffraction patterns of \CePrBx\ measured in the paramagnetic phase for (from top to bottom) $x=$ 0.2 ($T=$ 9.9 K), 0.4 (10.0 K), and 0.8 (8.0 K).}
\end{figure}

All compounds were first measured in the paramagnetic phase around 10 K, a few degrees above the Néel temperature. The diffraction patterns were refined in the $Pm\bar{3}m$ crystal structure, as shown in Fig.~\ref{Ce0.8nuc} for $x = 0.8$. The resulting parameters are listed in Table~\ref{crysparam}. The boron coherent length was also refined and the values obtained are in good agreement with those calculated from the nominal $^{11}$B isotope content. A faint, broad, somewhat asymmetric contribution appears in the feet of the Bragg peaks (Fig.~\ref{paraphase}), denoting the existence of short-range order, with an isotropic correlation length of the order of 100 \AA. A small fraction of a second phase gives rise to weak peaks above $2\theta \approx 50\dg$ (Fig.~\ref{paraphase}), but these peaks do not hinder the analysis of the magnetic signal located mainly at lower angles. More intriguing is the existence of a very weak extra contribution, with a markedly asymmetric, 2D-like, profile, near $2\theta \approx 13\dg$. This feature was observed consistently for the 3 compositions studied, but its origin is unknown.

Single crystals of Ce$_{0.8}$Pr$_{0.2}$$^{11}$B$_6$\ and Ce$_{0.7}$Pr$_{0.3}$$^{11}$B$_6$\ were grown by a floating-zone technique in pressurized, high-purity Ar gas, following the method described in Ref.~\onlinecite{Kunii'88}. Single crystal diffraction (SCD) measurements were performed on samples with volumes on the order of 0.1 cm$^{3}$, using the two-axis lifting detector neutron diffractometer 6T2 (LLB) at an incident wavelength $\lambda = 2.354$~\AA. Second-order contamination was suppressed by a PG filter inserted on the incident beam. A Soller collimator was placed before the counter to both reduce background and improve \V{Q} resolution. The samples were cooled down to a minimum temperature of 1.5 K in an Oxford Instruments cryomagnet. An Air Liquide \hbox{$^3$He-$^4$He} dilution insert was further utilized for measurements on \CePrB{0.8}{0.2}\ down to 100 mK. High-field results for  \CePrB{0.7}{0.3}\ were discussed briefly in Ref.~\onlinecite{Mignot'06}, and a more complete account for the two compositions $x = 0.7$ and 0.8 will be reported in a forthcoming paper. Here we focus on the results obtained in zero-field for both powder and single-crystal materials.


\section{\label{sec:magpwdr} Powder diffraction results}

 \begin{figure}  
 	\includegraphics [width=0.55\columnwidth, angle=-90] {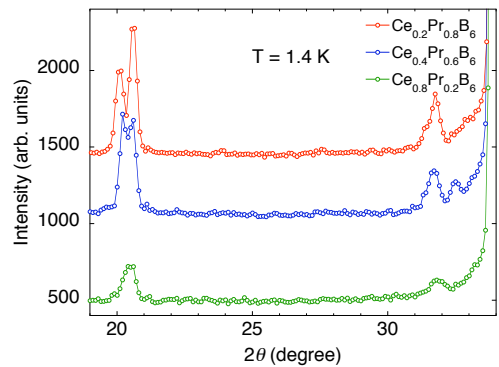}
\caption{\label{allconc_tmin} (Color online) Evolution of the low-angle magnetic satellites at $T=$ 1.4 K in \CePrBx\ as a function of composition. Intensities for each sample have been normalized with respect to the nuclear Bragg peaks (for clarity, each curve has been shifted upward by 500 relative to the previous one).}
\end{figure}

 \begin{figure}  
	\includegraphics [width=0.55\columnwidth, angle=-90] {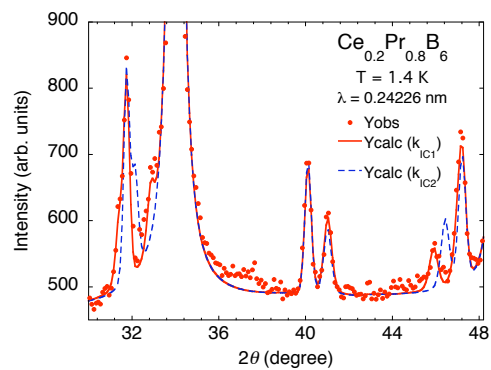}
\caption{\label{compxy} (Color online) Diffraction pattern measured at $T=$ 1.4 K for $x=$ 0.2: comparison of refinements taking the magnetic wave vector to be either $\mathbf{k}_{\mathrm{IC1}}$ or $\mathbf{k}_{\mathrm{IC2}}$.}
\end{figure}

For these measurements, the samples were first cooled down to $T_{\mathrm{min}} \approx 1.4$ K, then heated by steps of 0.5 K to $T \approx 10~\mathrm{K} > \TN$. The low-angle part of the diffraction patterns measured at $T_{\mathrm{min}}$ is presented in Fig.~\ref{allconc_tmin}. For $x = 0.2$ and 0.4, one notes the existence of two sets of magnetic satellites. Their positions can be indexed using the same type of magnetic \V{k} vectors previously identified\cite{Burlet'88} in the incommensurate $[\mathbf{k}_{\mathrm{IC1}}=(1/4-\delta,1/4,1/2)]$ and commensurate $[\mathbf{k}_{C}=(1/4,1/4,1/2)]$ ordered phases of pure PrB$_6$ (hereafter, the different magnetic phases are termed according to Kobayashi \etal\cite{Kobayashi'03}). The Bragg peaks associated with the two components are close together, but still clearly separated within experimental resolution. Refinements yield values of $\delta$ equal to 0.0394(6) and 0.0278(5) for $x = 0.2$ and 0.4, respectively, to be compared to $\delta \approx 0.05$ for pure PrB$_6$ in Ref.~\onlinecite{Burlet'88}. Ce substitution thus produces a steady, and rather pronounced, reduction of the incommensurability in the Pr-rich compounds. An alternative refinement was also tested using a wave vector $\mathbf{k}_{\mathrm{IC2}}=(1/4-\delta ',1/4-\delta ',1/2)$. The doublet of peaks at the lowest scattering angle (000$^{\pm}$ satellites) is again reproduced satisfactorily, yielding $\delta '= 0.0196(3)$ for $x=0.2$. On the other hand the model fails to account for the groups of satellites located at $2\theta \approx 31.5$\dg\ and 46.5\dg\ (Fig.~\ref{compxy}), and the magnetic $R$-factors for both the incommensurate and commensurate components degrade significantly (19.5 and 15, respectively) as compared to the previous refinement using $\mathbf{k}_{\mathrm{IC1}}$ (16 and 12.5). The first model is thus clearly favored for this composition, and the same holds true for $x = 0.4$.

For $x = 0.8$, the magnetic peak intensities displayed in Fig.~\ref{allconc_tmin} are strongly reduced, and the double-peak structure is no longer resolved, even for the strongest satellite near $2\theta = 20.6\dg$. However, the shape (and width) of the peak, as well as its temperature dependence (see below), still point to the existence of two distinct, C and IC, components. As above, all refinements were performed assuming either $\mathbf{k}_{\mathrm{IC1}}=(1/4-\delta,1/4,1/2)$ or $\mathbf{k}_{\mathrm{IC2}}=(1/4-\delta ',1/4-\delta ',1/2)$, but here the experimental accuracy was not sufficient to discriminate between the two solutions. The values obtained for $\delta$ and $\delta '$ (0.021(3) and 0.011(2), respectively, for $T = 1.4$ K) verify $\delta '\approx \delta/2$, as expected if the refinement is dominated by the 000$^\pm$ satellite, which is insensitive to the type of $\mathbf{k}$ vector chosen. In the following, we retain the latter solution (incommensurate component oriented along $\langle110\rangle$), which is supported by the single-crystal results presented in Section~\ref{sssec:magce:sc:0.8} below. Extra satellites associated with the second type of wave vector, $\mathbf{k}'_{C}=(1/4,1/4,0)$, indicative of the so-called ``double-$k$--$k'$'' structure reported in Ref.~\onlinecite{Effantin'85} for phase III of pure CeB$_{6}$, could not be detected in the present NPD measurement. It will be shown in the next section that these extra reflections actually exist, but with intensities too weak to be detected in the powder data. 

 \begin{figure}  
	\includegraphics [angle=-90, width=0.85\columnwidth] {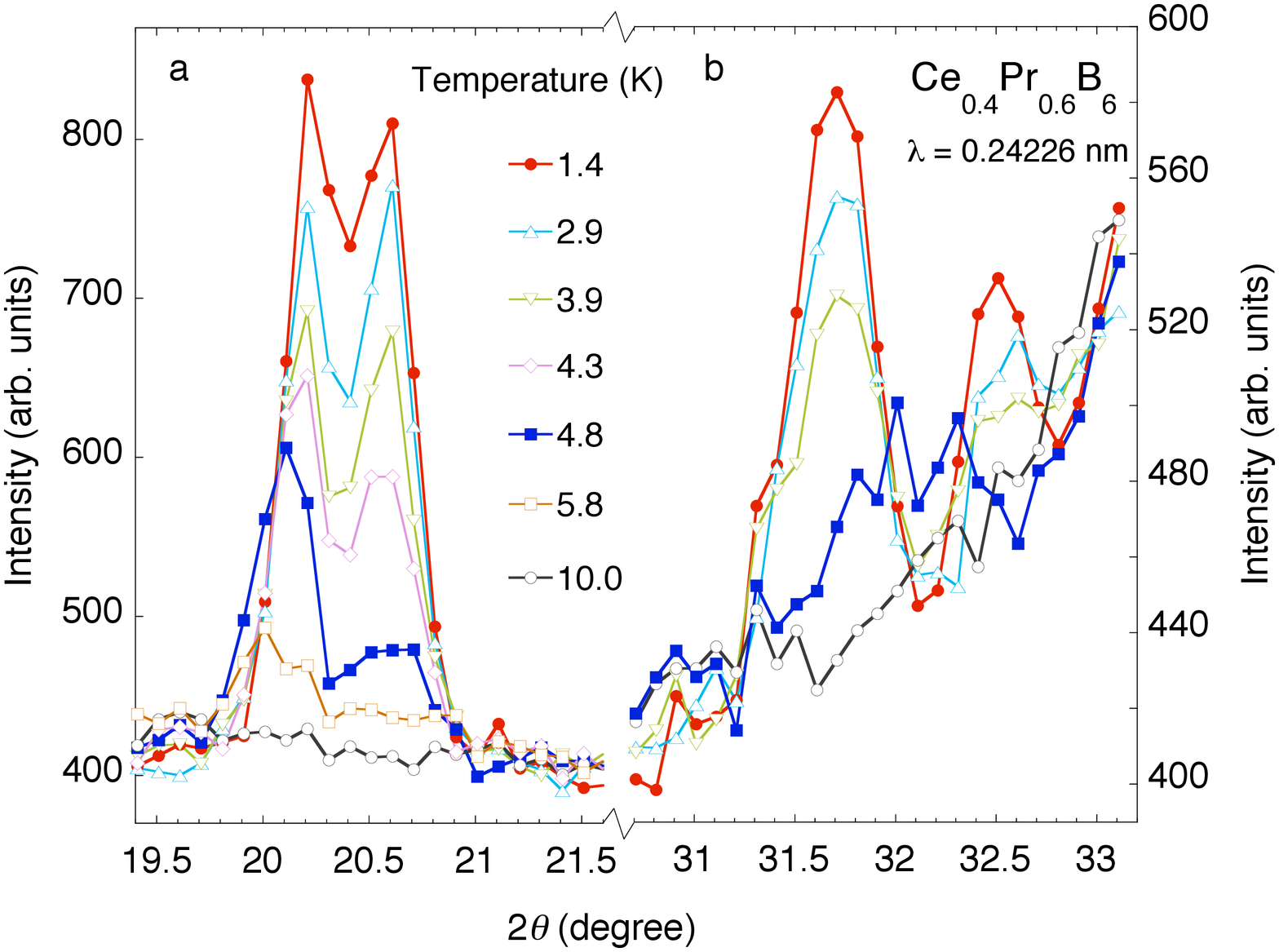}
	\par \vspace{12pt}
	\includegraphics [angle=-90, width=0.425\columnwidth] {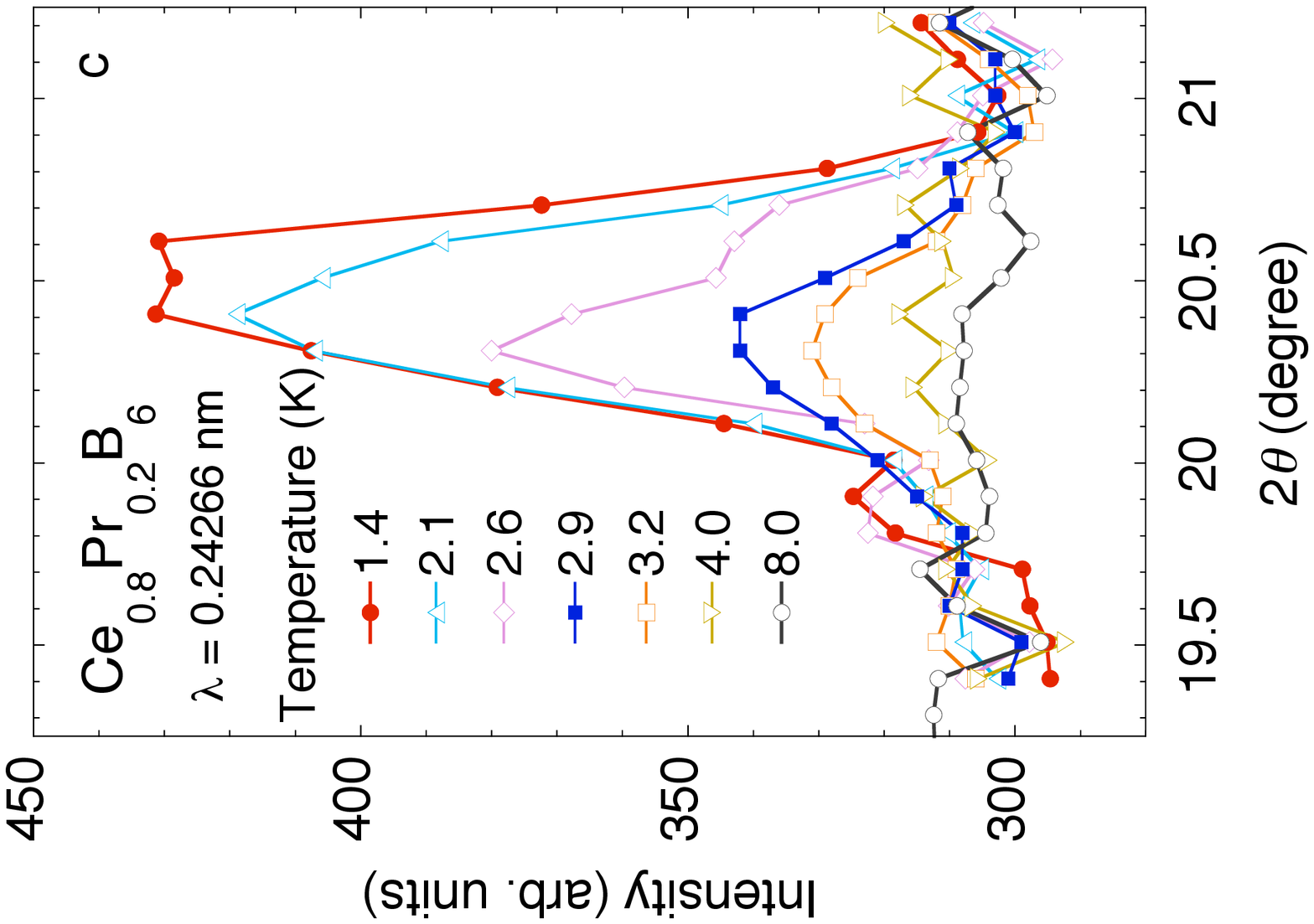}
\caption{\label{Ce0.4-0.8_alltemps} (Color online) Temperature dependence of the lower-angle magnetic satellites (a): 000$^{\pm}$, and (b): 101$^{-}$, 011$^{-}$ in \CePrB{0.4}{0.6}; (c): 000$^{\pm}$ in \CePrB{0.8}{0.2}. Data for some intermediate temperatures have been omitted for clarity}
\end{figure}

 \begin{figure}  
	\includegraphics [width=1\columnwidth, angle=0] {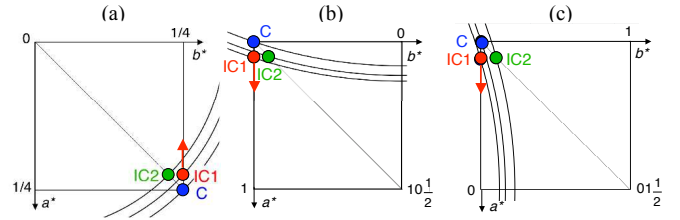}
\caption{\label{qspace} (Color online) Schematic representation (incommensurate components not to scale) of 3 different regions of the $l = 1/2$ plane in reciprocal space, showing the position of the Bragg satellites associated with $\mathbf{k}_{C}$, $\mathbf{k}_{\mathrm{IC1}}$, and $\mathbf{k}_{\mathrm{IC2}}$. Panel (a) (000$^{+}$ reflections) corresponds to the scattering angle range plotted in Fig.~\ref{Ce0.4-0.8_alltemps}(a), panels (b) (101$^{-}$ reflections) and (c) (011$^{-}$ reflections) to that plotted in Fig.~\ref{Ce0.4-0.8_alltemps}(b).}
\end{figure}

 \begin{figure} 
	\includegraphics [width=0.65\columnwidth, angle=0] {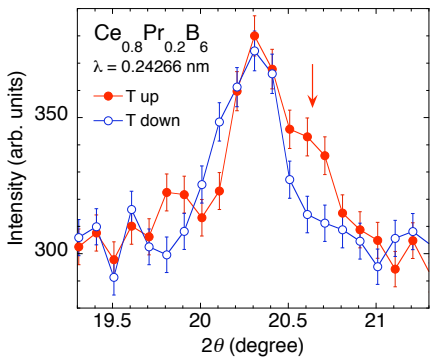}
\caption{\label{T2.58updown} (Color online) Comparison of the lower-angle satellites at $T = 2.6$ K measured upon heating and cooling. The arrow indicates the position of the commensurate peak.}
\end{figure}

 The temperature evolution of the lower two groups of satellites displayed in Figs.~\ref{Ce0.4-0.8_alltemps}(a) and \ref{Ce0.4-0.8_alltemps}(b) for \CePrB{0.4}{0.6}\  indicates that the IC and C components decrease simultaneously on heating up to $\approx 4$ K. Above 5 K, the intensity of the commensurate component drops more steeply whereas the incommensurate satellite 000$^{\pm}$ in Fig.~\ref{Ce0.4-0.8_alltemps}(a) moves to lower scattering angles. This shift, however, cannot be the result of an increase in $\delta$, as one might think at first [Fig.~\ref{qspace}(a)]: if it were the case, the well-defined 101$^{-}$ satellite, occurring on the right side of the C peak in Fig.~\ref{Ce0.4-0.8_alltemps}(b), should move to \textit{larger} scattering angles as depicted schematically by the arrows in Fig.~\ref{qspace}(b). Experimentally, we instead observe a transfer of intensity from the IC1 satellites at 31.5 and 32.5\dg\ to the intermediate region near 32.1\dg. This can be explained, as represented in Fig.~\ref{qspace}, by ascribing the latter signal to a new magnetic structure, still incommensurate, but with a wave vector of the form $(1/4-\delta',1/4-\delta',1/2)$. Our results thus support the existence, over a limited temperature range below \TN, of a distinct magnetic phase, whose incommensurate component is oriented along the [110] direction. It is worth noting that the existence of this phase, hereafter denoted IC2, was previously reported from bulk measurements\cite{Kim'06} for an intermediate composition $x = 0.3$. 

In \CePrB{0.8}{0.2} (Fig.~\ref{Ce0.4-0.8_alltemps}), the center of mass of the peak doublet shifts to lower angles with increasing temperature, indicating a reduction in the relative weight of the C component. No direct evidence is found, however, for the IC--C first-order phase transition at $T=1.7$ K observed by Nagai \etal , \cite{Nagai'08} and confirmed by the single-crystal results presented hereafter in Section \ref{sssec:magce:sc:0.8} for the same Ce concentration. However, the comparison of patterns measured at $T = 2.6$ K after heating from 1.4 K or cooling from the paramagnetic phase (Fig.~\ref{T2.58updown}) reveals a sizable difference in the magnitude of the commensurate component, which might be connected with the irreversibility at this transition.

 \begin{figure}  
	\includegraphics [width=0.75\columnwidth, angle=0] {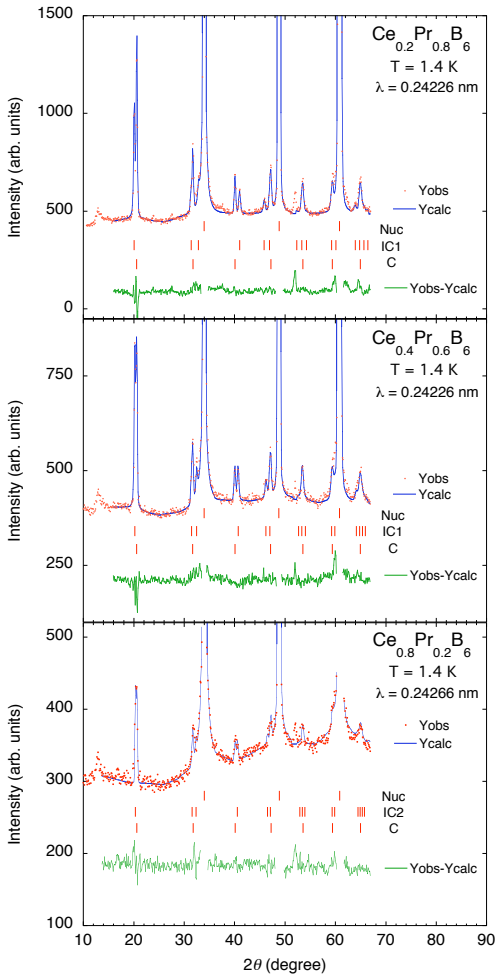}
\caption{\label{prf_T1.4} (Color online) Expanded view of the powder diffraction patterns measured at $T=$ 1.4 K for \CePrBx\ ($x=$ 0.2, 0.4 and 0.8), showing the refinement of the magnetic peaks.}
\end{figure}

\begin{table} [b]
\caption{\label{magmoments}Refined parameters $\delta$ and $R$ and average magnetic moments $m$ in \CePrBx\ at $T_{\mathrm{min}} = 1.4$ K calculated using the assumptions discussed in the text. Values for $R$ and $m$ are given in units of Bohr magnetons}
\begin{ruledtabular}
\begin{tabular}{c c c c c c c c c}
$x$ & $\delta [\delta']$ & $R_{C}$ & $R_{\mathrm{IC1}} [R_{\mathrm{IC2}}]$ & $m$\\
0.2 & 0.0394(6) & 1.37(6) & 1.24(7) & 1.85(9) \\
0.4 & 0.0278(5) &1.08(8) & 1.25(8) & 1.65(11) \\
0.8 & [0.011(2)] & 0.59(5) & [0.60(5)] & 0.84(7)\\
\end{tabular}
\end{ruledtabular}
\end{table}

In view of the strong similarities between the present results for $x=0.2$ and 0.4 and those reported earlier by Burlet \etal \cite{Burlet'88} for pure PrB$_{6}$, it is very likely that the structures associated with the C and IC1 magnetic components are of the same type here as in the latter compound, namely planar, noncollinear, \hbox{double-$k$} structures. The real-space description of the different structures given in Ref.~\onlinecite{Burlet'88} is based on the extra assumption that the moments at all Pr sites are equal, and can be expressed as
\begin{eqnarray}
\mathbf{m}_{C}(\mathbf{r}) & = & R_{C}\left[\cos(\mathbf{k}_{C}^{(1)}\cdot \mathbf{r})\hat{\mathbf{u}}_1 + \sin(\mathbf{k}_{C}^{(2)}\cdot \mathbf{r})\hat{\mathbf{u}}_2 \right], \nonumber \\
\mathbf{m}_{\mathrm{IC1}}(\mathbf{r}) & = & R_{\mathrm{IC1}} \left[\cos(\mathbf{k}_{\mathrm{IC1}}^{(1)}\cdot \mathbf{r})\hat{\mathbf{u}}_1 + \sin(\mathbf{k}_{\mathrm{IC1}}^{(2)}\cdot \mathbf{r})\hat{\mathbf{u}}_2 \right],
\label{eq_magstruc}
\end{eqnarray}
where
\begin{eqnarray}
\hat{\mathbf{u}}_1 & = & \frac{1}{\sqrt{2}}(1,-1,0), \nonumber \\
\hat{\mathbf{u}}_2 & = & \frac{1}{\sqrt{2}}(1,1,0)
\label{eq_unitvec}
\end{eqnarray}
are orthogonal unit vectors in the (001) plane. The (1) superscripts denote the \V{k} vectors defined above, and the (2) superscripts those obtained by changing the sign of the second component. 

Regarding phase IC2, evidence has been found very recently that it is indeed \hbox{single-$k$}.\cite{Mignot'up,Robert'up} This question, however, is irrelevant to the present zero-field study, since diffraction experiments performed without an external symmetry-breaking field cannot discriminate between \hbox{single-$k$} and \hbox{multi-$k$} structures.\cite{Rossat'87} For convenience, all intensity refinements were thus carried out for the \hbox{single-$k$} case, i.e. with the magnetic moment simply given by
\begin{eqnarray}
\mathbf{m}_{\lambda}(\mathbf{r}) & = & R_{\lambda} \cos(\mathbf{k}_{\lambda}^{(1)}\cdot \mathbf{r})\hat{\mathbf{u}}_1
\label{eq_magstruc1k}
\end{eqnarray}
with $\lambda = \{ \mathrm{C}, \mathrm{IC1}, \mathrm{IC2}\}$. In these calculations, each component was treated separately, i.e. assuming that its magnetic intensity originates from the entire sample volume. The corresponding refinements for $T=T_{\mathrm{min}}$ are shown in Fig.~\ref{prf_T1.4}, and the parameters from the fit are collected in Table~\ref{magmoments}. It can be noted that the C component is larger than the IC1 component for $x = 0.2$, but smaller for $x = 0.4$. This is in line with the suppression of the lock-in transition with increasing Ce content observed in bulk experiments.\cite{Kim'06}

 \begin{figure}  
	\includegraphics [angle=0,width=0.90\columnwidth] {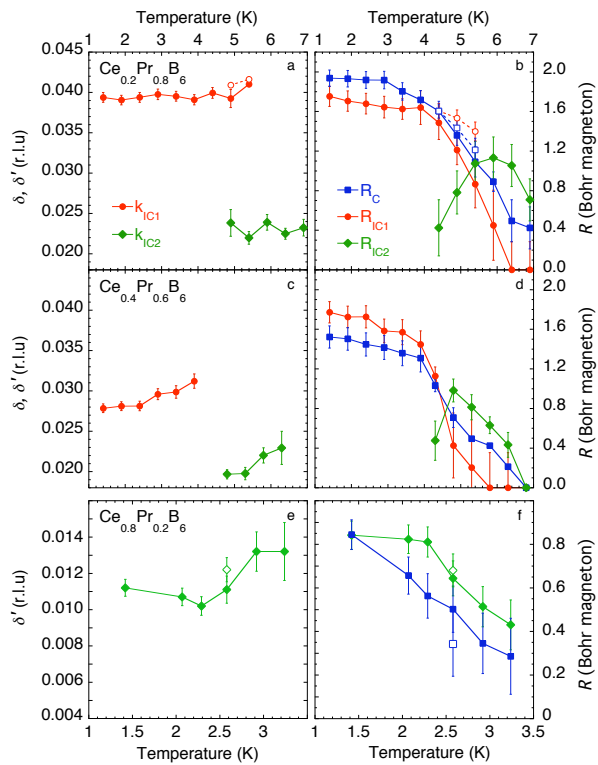}
\caption{\label{Tdpdces} (Color online) Temperature dependence of the incommensurate parameters $\delta$ and $\delta'$ (left) and of the magnetic Fourier components (right) derived from the refinements of the powder diffraction patterns for \CePrBx\ ($x=0.2$, 0.4, 0.8). Closed (open) symbols represent data measured upon heating (cooling).}
\end{figure}

The temperature dependences of the IC components of the \V{k} vectors and of the magnetic Fourier components $R$ derived from the refinements are summarized in Fig.~\ref{Tdpdces}. The changes in $\delta$ and $\delta'$ are almost negligible for $x = 0.2$, slightly more pronounced for $x = 0.4$. For $x = 0.8$, the effect is difficult to ascertain because of the overlap between the two satellites, but the observed increase in $\delta'$ above 2.6 K is consistent with the single-crystal data presented in Section \ref{sec:magscry}. Comparing panels (b) and (d) in Fig.~\ref{Tdpdces}, one notes a reduction of \TN\ by about 1 K (from $\approx 7.5$ to $\approx 6.5$ K) between $x = 0.2$ and 0.4, in fair agreement with bulk measurements.\cite{Kim'06} The accuracy of the present estimate is however limited because neutron intensities, being proportional to the square of the magnetic moment components, become quite weak on approaching \TN.\footnote{The fact that the values of \TN\ from the NPD experiments are systematically higher than those derived from bulk measurements suggests a contribution from short-range fluctuations.} Above 5 K, the refinements indicate that the IC2 phase dominates, but that the original $\mathbf{k}_{C}$ and $\mathbf{k}_{\mathrm{IC1}}$ components are still present. At temperatures closer to the transition, it becomes difficult to determine the relative fractions of the two IC phases because the refinement is dominated by the strongest satellites 000$^{\pm}$, which are compatible with both $\mathbf{k}_{\mathrm{IC1}}$ and $\mathbf{k}_{\mathrm{IC2}}$. 
For $x=0.8$ (Fig.~\ref{Tdpdces}(f), the C component is suppressed more rapidly on heating than the IC component, with most of the variation taking place below 2.5 K.

To estimate the magnitude of the average rare-earth magnetic moments, we will ascribe the coexistence of the C and IC1 signals to different regions of the sample, rather than to a rather improbable structure involving these Fourier components simultaneously. This view is supported by the fact that, in single-crystal samples, the latter structures occur in different temperature ranges and are separated by a magnetic phase transition. Since the relative volume fractions are unknown, it was  further assumed that the magnetic moments $m$ have the same value in the two phases. The resulting values are listed in Table \ref{magmoments} for $T_{\mathrm{min}} = 1.4$ K, together with the parameters derived from the refinements. For $x = 0.2$ and 0.4, the moments are rather large and extrapolate to \hbox{2.05 $\mub$} at $x=0$, as compared to $m_{\mathrm{Pr}} = 1.2 \pm 0.1 \mub$ obtained in previous single-crystal measurements for the C phase of pure PrB$_{6}$.\cite{Effantin'85t} On the other hand, this value agrees quite well with that (\hbox{2.0 $\mub$}) expected for the $\Gamma_{5}$ ground state of Pr$^{3+}$. For $x=0.8$, the refined moment is smaller than that extrapolated from the Pr-rich compositions  (\hbox{1.05 $\mub$}), and than the value (\hbox{1.57 $\mub$}) expected for the $\Gamma_{8}$ ground state of Ce$^{3+}$. This effect is likely due to Kondo fluctuations of the Ce ions. However, the moment reduction is much less pronounced than that derived from the single-crystal data of Refs.~\onlinecite{Effantin'85} and \onlinecite{Effantin'85t} ($m_{\mathrm{Ce}}=0.28 \pm 0.06$ $\mub$). The reason for this disagreement is presently unknown.

 \begin{figure}  
	\includegraphics [width=0.75\columnwidth, angle=0] {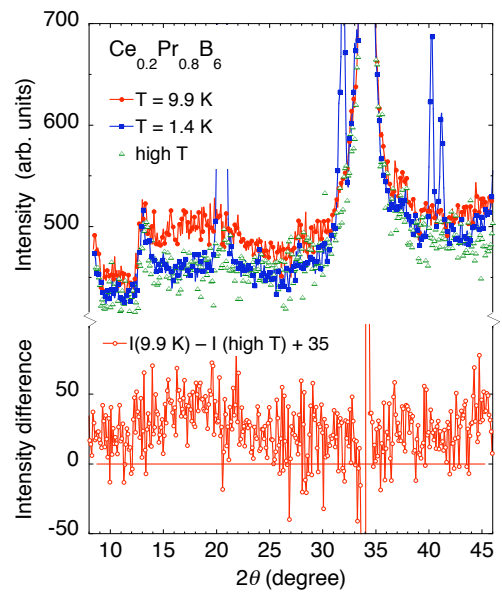}
\caption{\label{difference} (Color online) Diffuse magnetic signal  in \CePrB{0.2}{0.8} at different temperatures. The ``high $T$'' data (see text) have been corrected for a nuclear diffuse background of 35 counts, which does not exist at low temperature. This background was subtracted from the data at $T=9.9$ K to estimate the magnetic diffuse scattering (lower plot).}
\end{figure}
 
 \begin{figure}  [t] 
	\includegraphics [width=0.80\columnwidth, angle=0] {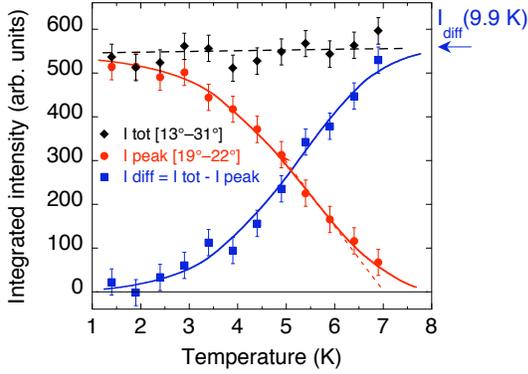}
\caption{\label{diffuse} (Color online) Temperature dependence of the integrated intensities of the magnetic Bragg (circles) and diffuse (squares) components in \CePrB{0.2}{0.8} at low-angles.}
\end{figure}

Finally, it is worth mentioning that, for $x = 0.2$ and 0.4, the magnetic background at low scattering angles increases substantially with increasing temperature. This effect is particularly visible in the region of the first magnetic satellites (Fig.~\ref{difference}). The signal was found to saturates near \TN, then to remain practically constant up to 10 K. The nuclear background was estimated by adding up data measured at several temperatures collected during the initial cool down of the sample below  150 K. Apart from an extra diffuse nuclear background of 35 counts (angle independent in this 2$\theta$ range), the signal between the Bragg peaks coincides with that measured at 1.4 K. This background was subtracted from the data measured at $T = 9.9$ K, yielding the difference plot shown in Fig.~\ref{difference}  for $x = 0.2$. One notes a broad maximum at $2\theta \approx 17\dg$, which can be ascribed to short-range magnetic correlations. This signal is centered at a significantly lower scattering angle than the magnetic Bragg reflections below \TN. Its width corresponds to a correlation length on the order of 15 \AA. Other maxima may exist at higher angles but the statistics on the difference signal is not sufficient to resolve them. 
Below \TN, the magnetic satellites grow at the expense of the short-range signal as demonstrated in Fig.~\ref{diffuse}, where the intensity integrated between 13 and 31 degrees in $2\theta$ is seen to exactly balance the growth of the Bragg peaks. The existence of magnetic fluctuations persisting up to about $3T_{N}$ in PrB$_{6}$ was detected previously by $^{11}$B NMR experiments.\cite{Takagi'85}


\section{\label{sec:magscry}Single-crystal diffraction results}       
\subsubsection{\label{sssec:magce:sc:0.8} $x = 0.8$}

\begin{figure} 
	\includegraphics [width=0.475\columnwidth, angle=-90] {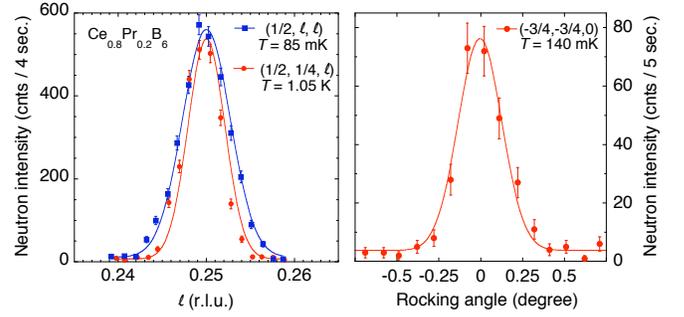}
\caption{\label{SX0.8_dilu} (Color online) Commensurate magnetic satellites in phase III of \CePrB{0.8}{0.2}, demonstrating the existence of the wave vectors $\mathbf{k}_{C}= (1/4, 1/4, 1/2)$ (left) and $\mathbf{k}'_{C}= (1/4, 1/4, 0)$ (right).}
\end{figure}

For this composition, a first-order transition is known to take place at $\TIII=1.7$ K.\cite{Nagai'08} Below the transition, the magnetic order is found to be of the same type as in phase III in CeB$_{6}$, namely a noncollinear $2k$-$k'$ commensurate structure described by the four wave vectors $\mathbf{k}_{C}^{(1,2)}=(1/4, \pm 1/4, 1/2)$ and $\mathbf{k}_{C}'^{(1,2)}=(1/4, \pm 1/4, 0)$. Magnetic peaks associated with the different Fourier components have been observed at 85 mK and 140 mK  (Fig.~\ref{SX0.8_dilu}), and no significant change occurs in the intensity of the $\mathbf{k}_{C}$ satellite up to 1.05 K (see left frame). It can be noted that the relative magnitude of the $\mathbf{k}'_{C}$ peaks is strongly reduced in comparison with CeB$_{6}$,\cite{Effantin'85} which explains why it was not detected in the powder diffractograms (Section \ref{sec:magpwdr}). In the latter compound, the observation of two different wave vectors $\mathbf{k}_{C}$ and $\mathbf{k}'_{C}$ has been shown\cite{Effantin'85} to reflect the existence of two rare-earth sublattices in the (1/2, 1/2, 1/2) AFQ phase, which precludes a simple AF stacking of the 2$k$ planes of the type existing in PrB$_{6}$. This weakening of the $\mathbf{k}'_{C}$ component therefore points to a disruption of the AFQ order associated with the Ce sites as a result of Pr substitution.

 \begin{figure} 
	\includegraphics [width=0.5\columnwidth, angle=-90] {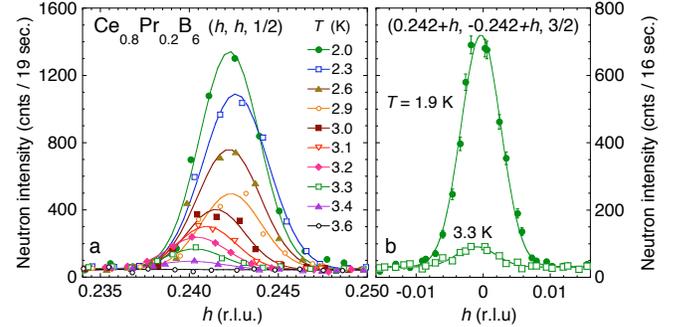}
\caption{\label{SX0.8_IC2} (Color online)  Temperature dependence of two incommensurate magnetic peaks, $[1/4-\delta']\:[1/4-\delta'] \: 1/2$ and $[1/4-\delta'] \: [-1/4+\delta'] \: 3/2$ in \CePrB{0.8}{0.2}. Scans were performed along orthogonal directions to emphasize (b) the existence of a single satellite  (IC2 phase) from 1.9 to 3.3 K, and (a) the variation of the  $\delta'$ parameter.}
\end{figure}

 \begin{figure} 
	\includegraphics [width=0.70\columnwidth, angle=0] {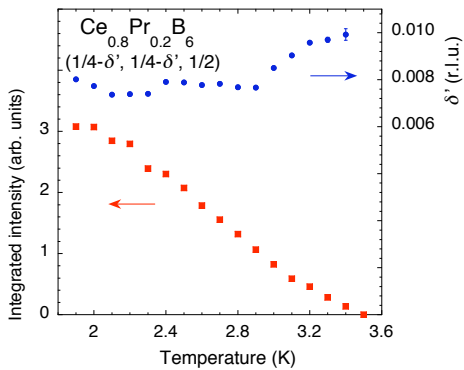}
\caption{\label{SX0.8_delta(T)} (Color online) Temperature dependences of the incommensurate parameter $\delta'$ (upper part) and of the integrated intensity of the magnetic peaks shown in Fig.~\ref{SX0.8_IC2}(a) (lower part) for \CePrB{0.8}{0.2}.}
\end{figure}

Above \TIII, the structure becomes incommensurate (the variation of the magnetic signal could not be traced across the transition because of temperature instabilities). At $T=2.0$ K, i.e.\ just above the transition to phase III, satellites are observed close to \V{Q} = $(1/4, \pm 1/4, 1/2)$ and $(1/4, \pm 1/4, 3/2)$ in agreement with the powder measurements. Here, however, the data unambiguously show that the structure is IC2-type, with the magnetic wave vector $\mathbf{k}'_{\mathrm{inc}}=(1/4-\delta',1/4-\delta',1/2)$, since only one peak is observed in the $(\xi, \xi, 0)$ scan through $\mathbf{Q} = (1/4-\delta', -1/4+\delta', 3/2)$  shown in Fig.~\ref{SX0.8_IC2}(b). The value of the incommensurate parameter $\delta'$, derived from the position of the Bragg peak in the left frame, is $1/4-0.242=0.008(1)$ at $T=2$ K, in fair agreement with the value of 0.011(2) reported above for powder. With increasing temperature, the peak position remains unchanged up to 2.9 K, then shifts gradually to lower $h$ values. Accordingly, the incommensurate parameter $\delta'$ increases from 0.008 to about 0.010 near $\TN = 3.5 \pm 0.05$ K (Fig.~\ref{SX0.8_delta(T)}). This evolution is quite similar to that displayed in Fig.~\ref{Tdpdces}(e), confirming that the effect was not an artifact of the Rietveld refinement. We note that the form of the wave vector remains $\mathbf{k}'_{\mathrm{inc}}=(1/4-\delta',1/4-\delta',1/2)$ up to the Néel temperature, as evidenced by the single satellite still found at $T = 3.3$ K $\lesssim \TN$ (right frame in Fig.~\ref{SX0.8_IC2}).  The change occurring above 2.9 K may correspond to the crossing of the AFQ phase boundary determined from bulk experiments\cite{Nagai'08} (see Fig.~\ref{phasediag}). The important point here is that the IC2 order is found to persist all the way down to \TIII\ implying that, in the phase diagram of Ref.~\onlinecite{Nagai'08}, an additional transition line must exist at finite field between this magnetic phase and the purely AFQ region (phase II), similar to that detected under pressure by the same authors. Contrary to the powder data (Section \ref{sec:magpwdr}), no magnetic signal corresponding to the commensurate wave vector $\mathbf{k}_{\mathrm{com}}$ was observed on the single crystal in the temperature range 2 K $\le T \le \TN$. Possible reasons for this difference will be discussed in Section \ref{sec:discussion}.

\subsubsection{\label{sssec:magce:sc:0.7} ${x = 0.7}$}

 \begin{figure} 
	\includegraphics [width=0.7\columnwidth, angle=0] {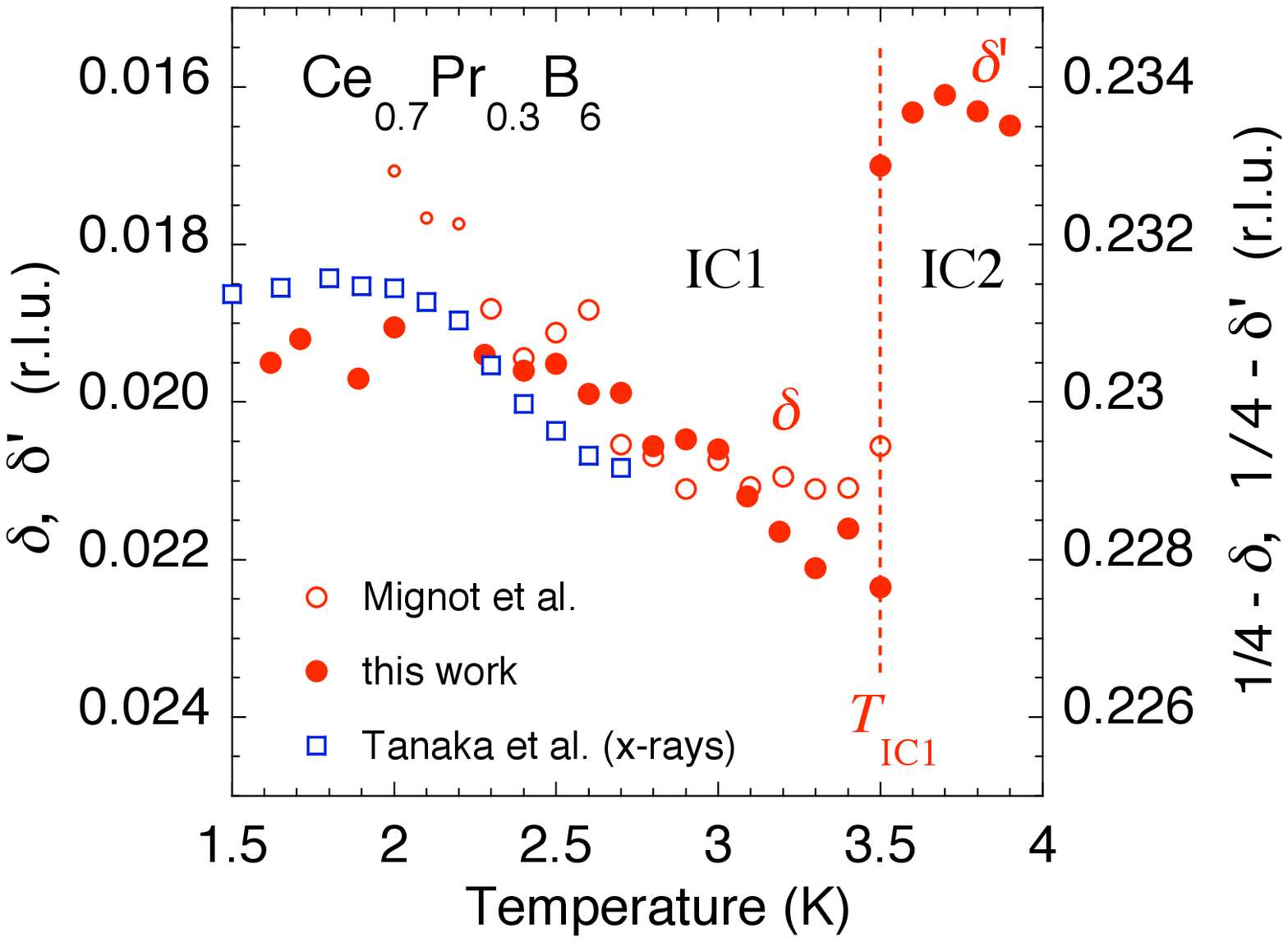}
\caption{\label{SX0.8_fcnT} (Color online) Temperature dependence of the incommensurate parameters $\delta$ and $\delta'$ in phases IC1 and IC2 of \CePrB{0.7}{0.3.}. Circles: neutron data from Ref.~\onlinecite{Mignot'06} and this work; squares: x-ray data from Ref.~\onlinecite{Tanaka'06}.}
\end{figure}

 \begin{figure} 
	\includegraphics [width=0.95\columnwidth, angle=0] {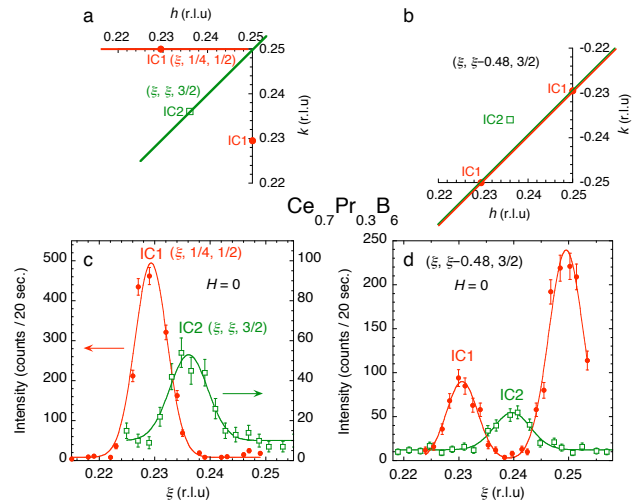}
\caption{\label{SX0.7_IC1-2} (Color online) Scans through magnetic satellites in phases IC1 ($T = 2$ K [c] and 1.7~K [d]) and IC2 ($T = 3.8$~K) of \CePrB{0.7}{0.3}: (left) along the direction of the incommensurate component; (right) perpendicular to it. The plots demonstrate the existence of two satellites in phase IC1 and a single satellite in phase IC2. The directions of the scans are depicted in the upper frames; although the $(\xi, \xi-0.48, 3/2)$ scan does not go exactly through the position of the IC2 satellite, the effect on the peak position is not significant within experimental resolution.}
\end{figure}

For the composition $x=0.7$, the results reported in Ref.~\onlinecite{Mignot'06} indicated that the magnetic structure remains incommensurate, with a wave vector $\mathbf{k}_{\mathrm{IC1}} = (1/4-\delta, 1/4, 1/2)$, down to $T=1.7$ K. This result is confirmed by the present measurements. However, the temperature variation of  $\delta$ is found to saturate below 2.3 K, rather than increasing further as reported previously.\cite{Mignot'06} As displayed in Fig.~\ref{SX0.8_fcnT}, a moderate, continuous decrease of $\delta$ takes place on heating up to the first-order transition at $T_{\mathrm{IC1}}= 3.5$ K. The nature of this transition has been studied in greater detail and it turns out that it corresponds to a change in the magnetic structure from IC1 to IC2. This effect was missed in the previous measurements because of instrumental resolution effects: in a scan along $(h,0,0)$ going through the position of the IC1 $(1/4-\delta, 1/4, 1/2)$ satellite, a peak is always observed, even after the structure changes to IC2, whose position does not represent the true value of the incommensurate wave vector. Unambiguous evidence for the distinct characters of the magnetic phases below and above $T_{\mathrm{IC1}}$ comes from the observation of two or one satellites, respectively, in the scans of Fig.~\ref{SX0.7_IC1-2}(d), as illustrated in frames (a) and (b). In phase IC2, no significant variation of $\delta '$ is observed up to the N\'eel temperature.

In summary, the main difference between the two Ce-rich compositions is the appearance of the IC2 phase for $x=0.7$, as well as the suppression of the commensurate phase III reminiscent of pure CeB$_{6}$. The existence of a finite \TQ\ in zero field is not clearly established for that composition, although it can be suggested from the observation of an $(1/2,1/2,1/2)$ component in the x-ray experiments at the lowest temperature ($T=1.5$ K).\cite{Tanaka'06}


\section{\label{sec:discussion}Discussion}

 \begin{figure} 
	\includegraphics [width=0.65\columnwidth, angle=-90] {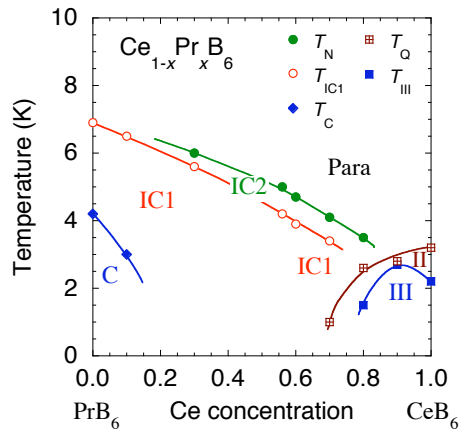}
\caption{\label{phasediag} (Color online) Composition--temperature phase diagram of \CePrBx. Transition temperatures are taken from Refs.~\onlinecite{Kim'06} and \onlinecite{Nagai'08} ($x=0.8$), and the labels for the different phases refer to the structures derived in this work.}
\end{figure}

A phase diagram summarizing the different magnetic phases investigated in this work is presented in Fig.~\ref{phasediag}.  In the Pr-rich composition range, the main changes produced by Ce substitution are i) a slow decrease of the Néel temperature, ii) the appearance, in an intermediate temperature interval between $T_{\mathrm{IC1}}$ and \TN, of a new magnetic phase with the IC component of its \V{k}\ vector oriented along a twofold axis, iii) in phase IC1: a rather pronounced reduction of the incommensurability parameter $\delta$, and iv) a moderate suppression of the C component. Points i) and ii) are in agreement with the composition--temperature phase diagram of Ref.~\onlinecite{Kim'06}. The new phase IC2 can be related to the high-field properties of pure PrB$_{6}$, in which a similar phase (also denoted IC2) exists \cite{Kobayashi'01} above $\approx 7$ T for $H \parallel \langle 111 \rangle$. That phase was found to occur at lower fields upon La dilution,\cite{Sera'05} and to exhibit the same type of IC wave vector, $(1/4-\delta',1/4-\delta',1/2)$, as in the present case.\cite{Robert'up} It is therefore likely that Ce substitution produces a dilution effect, similar to that of La, which eventually stabilizes phase IC2 all the way down to $H=0$.

Regarding the C component, the situation is rather confused. Whereas the phase diagram in Fig.~\ref{phasediag} suggests that the C phase is strongly suppressed by Ce substitution, with \TC\ dropping to zero above $x \gtrsim 0.1$, evidence for a C component is seen in the NPD patterns for both compositions $x = 0.2$ and 0.4, with just a minor decrease in its magnitude at the higher concentration. Moreover, there is no indication of a lock-in transition and the reflections associated with the C wave vector coexist with the IC1 or IC2 component practically up to the transition into the paramagnetic state. Tests performed on PrB$_{6}$ powder showed that, even in that pure compound, the C component can be traced at all temperatures below \TN, in contrast to the sharp lock-in transition observed on single crystals.\cite{Burlet'88,Kobayashi'01} The reason for this discrepancy is presently unclear. It can be noted that the problem is specific to the commensurate component and the transition at \TC, whereas consistent results are obtained for \TN\ and $T_{\mathrm{IC1}}$. Temperature irreversibility may play a role, since all measurements were performed after cooling the sample down to $T_{\mathrm{min}}$, but the sequence of \textit{two} different IC phases observed as a function of temperature, both coexisting with the C component, is difficult to reconcile with this view. One may suspect that the delicate balance of interactions controlling the competition between the C and IC phases is easily disturbed by defects or inhomogeneities, which are more likely to occur in the case of large powder samples. However, even for PrB$_{6}$ single crystals, differences have been found between samples of the same origin, as in Ref.~\onlinecite{Sera'96} where \TC\ is clearly visible in the thermal conductivity, but not in the electrical resistivity. Further investigations are needed in order to solve this problem. 

On the Ce-rich side, one notes a strong reduction in the intensity of the magnetic signal for $x=0.8$, accompanied by a further decrease of the incommensurability parameter. From the SCD data, it is clear that the structure prevailing below \TN\ down to the lower transition temperature \TIII\ is of IC2 type. The fact that this magnetic state characteristic of PrB$_{6}$ can survive for 80\% Ce concentration is quite remarkable and implies that Pr--Pr interactions are dominant over a wide composition range. On the other hand, the observation of \hbox{$\mathbf{k}'$-type} $(1/4,1/4,0)$ satellites in phase III is the fingerprint of Ce order occurring below \TQ, as in pure CeB$_{6}$. \CePrB{0.8}{0.2} thus represents a crossover situation where Ce and Pr features coexist.  By comparing the neutron data with the phase diagram of Ref.~\onlinecite{Nagai'08}, one is led to the conclusion that the crossing of the AFQ line at $\TQ \approx 2.9$ K does not produce any notable change in the magnetic satellites, and that the phases denoted IC2 and IC2' are thus associated with identical, or very similar, magnetic structures.

For $x=0.7$, the AFQ ordering temperature, if any, is located below 2 K. This suppression of \TQ\ is ascribed to the competition with the IC magnetic order, which is enhanced for this composition in comparison with $x=0.8$.\cite{Nagai'08} At higher temperature, two distinct IC magnetic phases occur successively on cooling. As noted above, each of them is closely related to a counterpart existing in pure PrB$_{6}$, suggesting that Ce plays only a minor role here in determining the type of order. 
In Ref.~\onlinecite{Kishimoto'05}, it was argued from magnetization data measured in fields as low as 0.1 T, that the anisotropy in phase IC2 is weak and that a domain with the ``$\chi_{\perp}$'' configuration is realized. The IC2--IC1 transition can be traced back to the growing effect of higher-order terms in the expansion of the free energy, due to QP interactions, as temperature decreases. In PrB$_{6}$, the first-order character of the transition at \TN\ has been argued to result from a competition between (anisotropic) exchange and QP interactions.\cite{Burlet'88} In the present case, the transition at $T_{\mathrm{IC1}}$ is clearly first-order, whereas that at $T_{N}$ looks second-orderlike. This further supports the idea that QP interactions do not play a major role in phase IC2 but become significant in phase IC1. AFQ fluctuations may be significant even at temperatures higher than the extrapolated $T_{Q}$ at $H = 0$ and, accordingly, affect the magnetic order setting in below $T_{\mathrm{IC1}}$.

It is particularly interesting, in this connection, to compare the neutron results for $x=0.7$ with nonresonant XRD data obtained on the same compound.\cite{Tanaka'06} In that study, satellites were observed at incommensurate positions in phase IC1 at $T = 1.5$ K. Their intensities decrease gradually on heating and vanish below approximately 3 K. No superstructure was then detected in phase IC2. The important point is that a close relationship exists between magnetic and charge ordering in this regime since the wave vector derived from the XRD peaks is precisely twice that of the ICM order observed by neutrons: 

\begin{eqnarray}
\mathbf{k}_{Q} & = & (0.462, 1/2, 0) \nonumber \\
	& = & 2(0.231, 1/4, 1/2) - \bm{\tau}_{001} \approx 2 \mathbf{k}_{M} - \bm{\tau}_{001}
\label{eq:k2k}
\end{eqnarray}

This observation is further substantiated by the good agreement seen in Fig.~\ref{SX0.8_fcnT} between the temperature dependence of the incommensurate component $\delta$ derived from the x-ray results using Eqn.~\ref{eq:k2k} and that measured by neutron diffraction. In Ref.~\onlinecite{Tanaka'06}, Tanaka \etal\ analyzed the wave vector dependence of the intensities of different IC satellites in terms of the structure factors for quadrupole and hexadecapole moments, showing that their data imply an incommensurate multipole (\ICMu) ordered state in phase IC1. Whereas there exists a clear interplay between magnetic and multipole orders occurring in this phase, the exact nature of the primary order parameter remains an open question. At the minimum temperature of 1.5 K, additional x-ray peaks are observed at positions $(n/2, n/2, n/2)$, reflecting an extra antiferro-multipolar component, which the authors identify with the ground state of the system. It is tempting to ascribe its appearance to the crossing of the $T_{Q}$ transition line, as shown in the phase diagram of Kishimoto \etal \cite{Kishimoto'05} Unfortunately, this range of temperature was not accessible in our neutron experiments. In view of the striking similarity between the magnetic properties of phase IC1 for the present composition and in PrB$_{6}$, it would be worthwhile to perform XRD experiments in order to see if a similar \ICMu\ order also exists in the latter compound.
\\


\section{\label{sec:conclusion}Conclusion}

The work presented in this report provides detailed information on the nature of the different ordered states occurring in \CePrBx\ solid solutions as a function of the rare-earth composition. This is important from the point of view of multipole interactions because $O_{xy}$-type quadrupolar couplings are known to play a major role throughout the series, but the long-range ordered AFQ state exists only on the Ce-rich side, where octupole interactions are significant. As a result, quite different magnetic phases are stabilized in one or the other case. The $2k$-$k'$ phase III is only observed at temperatures below the AFQ transition and, in cases when magnetic order sets in above this transition, for $x=0.7$ and 0.8, the corresponding structure is incommensurate and clearly dominated by Pr-Pr interactions. The comparison of neutron and x-ray diffraction results proved quite useful in the case of the $x=0.7$ composition since it revealed a perfect matching between the incommensurate multipole and magnetic wave vectors, which confirms that both phenomena are intimately related. This approach should be extended to other concentrations, including pure PrB$_{6}$. The sequence of two incommensurate phases, IC2 (single-$k$) and IC1 (double-$k$), with decreasing temperature is the dominant feature for a wide range of concentrations ($0.2 \leq x \leq 0.7$). The fact that the two transition lines remain nearly parallel as a function of $x$ suggests that the double-$k$ structure becomes favorable when the magnetic order parameter reaches a threshold value below \TN. Measurements in external magnetic fields are essential, both to identify multi-$k$ structures from domain repopulation effects and to study the competition between different types of order parameters occurring in these systems. Magnetic and transport measurements have shown that remarkably rich ($H$, $T$) phase diagrams exist depending on composition or applied pressure, but a number of phases have not yet been characterized. Experiments along this line are under way and will be reported in a forthcoming publication.

\begin{acknowledgments}
We are grateful to A.\ Gukasov, Ph.\ Boutrouille, J.-L.\ Meuriot, and Th.\ Robillard for help with the single-crystal experiments. We thank one of the Referees for useful suggestions on the presentation of the paper.
\end{acknowledgments}



\end{document}